# Light-emitting diode spherical packages: an equation for the light transmission efficiency

Ivan Moreno[1]*, David Bermúdez[2], Maximino Avendaño-Alejo[3]

[1] *Physics Faculty, Universidad Autónoma de Zacatecas, Zacatecas 98060, México*
[2] *Departamento de Física, Cinvestav, AP14-740, 07000 México D.F., Mexico.*
[3] *Universidad Nacional Autónoma de México, Centro de Ciencias Aplicadas y Desarrollo Tecnológico, Distrito Federal 04510, México.*
*Corresponding author: imoreno@fisica.uaz.edu.mx

Virtually all light-emitting diodes (LEDs) are encapsulated with a transparent epoxy or silicone-gel. In this paper we analyze the optical efficiency of spherical encapsulants. We develop a quasi-radiometric equation for the light transmission efficiency, which incorporates some ideas of Monte-Carlo ray tracing into the context of radiometry. The approach includes the extended source nature of the LED chip, and the chip radiance distribution. The equation is an explicit function of the size and the refractive index of the package, and also of several chip parameters such as shape, size, radiance, and location inside the package. To illustrate the use of this equation, we analyze several packaging configurations of practical interest; for example, a hemispherical dome with multiple chips, a flat encapsulation as a special case of the spherical package, and approximate calculations of an encapsulant with a photonic crystal LED or with a photonic quasi crystal LED. These calculations are compared with Monte-Carlo ray-tracing, giving almost identical values.

# 1. INTRODUCTION

Today there are many light sources, but LEDs are the ones that offer the most promising future, due to energetic efficiency, life-time, and controllability [1,2]. Consequently, extensive investigations are being conducted to increase the optical efficiency of LEDs. In this effort, the light transmission efficiency (LTE) of the package plays an important role [3-13].

LEDs are packaged because three major reasons: 1) to increase the optical efficiency, 2) to protect the die from environment, and 3) to shape the radiation pattern of light emission. This makes that virtually all light-emitting diodes (LEDs) are encapsulated with a transparent epoxy or silicone-gel. Various types of high power LED packages are used, but spherical and flat are the most popular. The use of spherical packages has rapidly spread, due to improvements in packaging methods [4-6], materials [6-8], and designs [9-12], which has reduced the fabrication cost and increased both the efficiency and lifetime [13].

A packaged LED is a light source encapsulated inside a special lens. When the minimum feature size of a light source is of the same order as that of the wavelength of light, a rigorous electromagnetic analysis is needed, e.g. for calculating the light extraction efficiency of photonic crystal LEDs [14-17]. If the minimum size is much greater than the wavelength, the nonimaging optics approach is used [18,19]. Therefore, the analysis of the LTE of an encapsulating lens is within the domain of geometrical optics and classical radiometry, and then is commonly studied by Monte Carlo ray tracing. In this process, useful assistance is provided by analytical methods. However, these are two-dimensional point source approaches [10,11,20], but the point source approximation is not accurate because of the finite dimensions of the chip with respect to encapsulating package [21]. As far as we know, a realistic analysis has not been reported before. In this work, we present a quasi-radiometric analysis of spherical packages to calculate the LTE in the tri-dimensional case, considering both the extended source nature of the LED chip and the chip radiation distribution. In addition, the LTE equation is calculated for several cases, and is verified with Monte-Carlo ray-trace simulation.

## 2. LIGHT TRANSMISSION EFFICIENCY

The LTE of the LED package is defined as the ratio of radiant flux exiting the package $\phi_{out}$ and the radiant flux emitted by the chip toward the package $\phi_{in}$.

$$\eta = \frac{\phi_{out}}{\phi_{in}} . \qquad (1)$$

In the materials for encapsulation there is no photon creation and the material can be chosen to be transparent in the emission spectrum of the LED chip. Therefore, $\eta$ is only affected by critical angle losses, Fresnel losses, reflection at the mirror cup, and absorption or recycling of light that returns to the chip [9-13,20,22-24]. We only consider the main factors, i.e., critical angle and Fresnel loss. It is to provide a practical conceptual framework and a useful tool to assist Monte Carlo ray tracing analysis. Therefore, the following equations give the minimum efficiency that may be obtained. On the other hand, the incorporation of multiple reflections and light recycling could be a topic for further research.

To calculate $\eta$, we need $\phi_{out}$ and $\phi_{in}$. Starting from the definition of radiance (W/m$^2$sr) or luminance (lm/m$^2$sr), the differential of emitted radiant flux is $d\phi = L_{in} d\Omega dA$; where $L_{in}$ is the radiance of the chip, $A$ is the projection of the radiative surface (in a given direction), and $\Omega$ is the solid angle in a given direction [25]. Thus, the flux radiated by the chip is

$$\phi_{in} = \int_{S_{in}} \int_{2\pi} L_{in} d\Omega dA \tag{2}$$

where the integration is done for every emitting point on the chip's surface $S_{in}$, and over the emitting hemisphere $\Omega = 2\pi$.

The radiometric theory states that the flux calculation starts in the radiance [25]. Therefore, for calculating the radiant flux exiting the package, the output radiance of the LED must be calculated. In general [25], the radiance change across the boundary between two homogeneous isotropic media is given by: $L_2 = (n_2/n_1)^2 \tau L_1$. Here $n_1$ and $n_2$ are the indices of refraction, and $\tau$ is the transmittance at the interface. Thus, considering that the light produced from the LED chip travels across the package boundary, the radiant flux exiting the package can be written as

$$\phi_{out} = \frac{n_{out}^2}{n_{in}^2} \int_{S_{out}} \int_{\Omega_{out}} L_{in} \tau \, d\Omega dA \tag{3}$$

where we have considered that $n_1 = n_{in}$ and $n_2 = n_{out}$ are the refractive indices of package and the output media, respectively. $S_{out}$ is the area of the spherical lens, and $\Omega_{out}$ is the solid angle for which the radiant flux comes out of the package. For example, for an LED with hemispheric encapsulating lens and a flat chip in its center, $\Omega_{out} = 0$ at the corner of the package because the projected area of the chip is zero there. However, trying to solve analytically this integral, one arrives to very complicated transcendental equations for the integration limits that define $\Omega_{out}=\Omega_{out}(A)$, thereby making the computation intractable. Recently, for calculating the radiation pattern of an encapsulated LED, we overcame this problem by carrying out the integral over the chip's paraxial image formed by the encapsulating lens [24]. But this approach was not suitable for obtaining $\eta$. In view of these challenges, we developed an alternative derivation of $\eta$ based on a hybrid approach.

## 3. QUASI-RADIOMETRIC APPROACH

We obtain $\phi_{out}$ without finding $\Omega_{out}(A)$ by a quasi-radiometric equation, which incorporates some ideas of Monte-Carlo ray tracing into the framework of radiometry. In Monte-Carlo ray tracing, a ray is basically a vector that simulates the radiation transfer. One defines an emitting source, from which a large number of light rays are randomly generated, and each of these rays contains a percentage of the total radiated flux. The rays are traced across the package, and are weighted by Fresnel losses and TIR. The number of rays (and their associated flux) in each segment of the package surface is counted to compute the output flux. The flux extracted from the package is added up to calculate the ratio of this and that from the chip, i.e. $\eta$. In the other hand, the radiometric approach leads to Eqs. (2) and (3).

In the following quasi-radiometric method, the percentage of flux emitted from every point of the chip that is transmitted for every solid angle is integrated over all the hemisphere. From radiometric theory, the flux portion emitted by a differential area of chip, *dS*, is

$$d\phi_{in} = L_{in} dA = L_{in}(\theta_p, \varphi_p, S)\cos\theta_p dS$$

The angles $\theta_p$ and $\varphi_p$ are spherical coordinates in a displaced coordinate system for every source point as is shown in Fig. 1 ($\theta_p$ is the polar angle, and $\varphi_p$ is the azimuthal angle). But how this flux portion is transmitted through the package?

In Monte-Carlo ray tracing, the total number of rays and their total amount of flux in a differential area of package is an element of differential flux. Because only the number of rays (and their associated fluxes) counts to compute the output flux, the package surface is not partitioned in "projected" differential areas as should be in a purely radiometric equation. Therefore, in the quasi-radiometric approach, the flux portion crossing a differential area of package, which is illuminated by a differential solid angle is

$$d\phi_{out} = L_{in}(\theta_p, \varphi_p, S)\tau(\alpha_{in})\cos\theta_p \sin\theta_p d\theta_p d\varphi_p dS .$$

The angle $\alpha_{in}$ is depicted in Fig. 1. The chip radiance is weighted by transmittance over the encapsulating sphere. Therefore, the total output flux, $\phi_{out}$, is

$$\phi_{out} = \int_{S_{in}} \int_0^{2\pi} \int_0^{\pi/2} L_{in}(\theta_p, \varphi_p, S)\tau(\alpha_{in})\cos\theta_p \sin\theta_p d\theta_p d\varphi_p dS , \qquad (4)$$

where the integration area $S_{in}$ is the chip emitting surface. Note that, from radiometric theory, a radiative transfer equation should include the projected area of the irradiated surface, i.e. $\cos\alpha_{in}$. If this cosine is incorporated in Eq. (4), the equation gives incorrect results because a formal radiometric approach leads to Eq. (3). However, as we will show in section 5, the quasi-radiometric equation and the Monte-Carlo ray-tracing simulation give almost identical results.

By using Eqs. (2) and (4), the LTE of the spherical package, $\eta$, is get

$$\eta = \frac{\int_{S_{in}} \int_0^{2\pi} \int_0^{\pi/2} L_{in}(\theta_p, \varphi_p, \boldsymbol{d}) \tau(\alpha_{in}) \cos\theta_p \sin\theta_p \, d\theta_p \, d\varphi_p \, dS}{\int_{S_{in}} \int_0^{2\pi} \int_0^{\pi/2} L_{in}(\theta_p, \varphi_p, \boldsymbol{d}) \cos\theta_p \sin\theta_p \, d\theta_p \, d\varphi_p \, dS} \quad , \tag{5}$$

and $\alpha_{in}$ is

$$\alpha_{in}(R, \theta_p, \varphi_p; \boldsymbol{d}) = \arccos\left(\frac{\boldsymbol{r} \cdot \boldsymbol{r}'}{|\boldsymbol{r}||\boldsymbol{r}'|}\right) = \arccos\left(\frac{\boldsymbol{r} \cdot \hat{\boldsymbol{r}}'}{|\boldsymbol{r}|}\right) \quad , \tag{6}$$

where $R$ is the radius of the package (Fig. 1), and we define

$$\hat{\boldsymbol{r}}' = \begin{pmatrix} \sin\varphi_p \cos\theta_p \\ \sin\varphi_p \sin\theta_p \\ \cos\varphi_p \end{pmatrix}, \quad \boldsymbol{d} = \begin{pmatrix} x \\ y \\ z \end{pmatrix}, \tag{7}$$

and from geometry we get

$$\boldsymbol{r} = \boldsymbol{d} + r'\hat{\boldsymbol{r}}', \quad r' = \left[\boldsymbol{d} \cdot \hat{\boldsymbol{r}}' | + \sqrt{(\boldsymbol{d} \cdot \hat{\boldsymbol{r}}')^2 + (R^2 - |\boldsymbol{d}|^2)}\right] \quad . \tag{8}$$

To evaluate the Eq. (5), we consider the transmittance for non-polarized light $\tau(\alpha_{in})$ with a step function $U$, as follows

$$\tau(\alpha_{in}) = T(\alpha_{in}) U(\alpha_{in} - \alpha c_{in}) \quad , \tag{9}$$

where, according to reference [26], $T$ is

$$T(\alpha_{in}) = \left(\frac{n_{out} \cos \alpha_{out}}{n_{in} \cos \alpha_{in}}\right) \frac{(t_P^2 + t_S^2)}{2} \quad . \tag{10}$$

Here the Fresnel transmission coefficients at the encapsulant surface for the electric fields parallel $t_P$ and perpendicular $t_S$ to the plane of incidence are given by

$$t_P = \frac{2 n_{in} \cos \alpha_{in}}{n_{in} \cos \alpha_{out} + n_{out} \cos \alpha_{in}}, \tag{11}$$

$$t_S = \frac{2 n_{in} \cos \alpha_{in}}{n_{in} \cos \alpha_{in} + n_{out} \cos \alpha_{out}}, \tag{12}$$

and the step function $U$ is

$$U(\alpha_{in} - \alpha c_{in}) = \begin{cases} 0 & \text{if } \alpha_{in} \geq \alpha c_{in} \\ 1 & \text{if } \alpha_{in} < \alpha c_{in} \end{cases}, \tag{13}$$

where the critical angle is

$$\alpha c_{in} = \arcsin\left(\frac{n_{out}}{n_{in}}\right) \quad . \tag{14}$$

In most cases only the angular intensity distribution $I_{in}(\theta_p, \varphi_p)$ [W/sr] may be available. In these cases, because the intensity $I_{in}$ is a spatially averaged angular distribution of radiance $L_{in}$, a good approximation for $\eta$ is proposed

$$\eta = \frac{\int\limits_{S_{in}}\int\limits_{0}^{2\pi}\int\limits_{0}^{\pi/2} I_{in}(\theta_p,\varphi_p) f(S) \tau(\alpha_{in}) \sin\theta_p \, d\theta_p \, d\varphi_p \, dS}{\int\limits_{S_{in}}\int\limits_{0}^{2\pi}\int\limits_{0}^{\pi/2} I_{in}(\theta_p,\varphi_p) f(S) \sin\theta_p \, d\theta_p \, d\varphi_p \, dS} \quad , \tag{15}$$

where $f(S)$ is a weighting function [27]. This introduces the position-dependent light strength across the chip surface. If $f(S)$ is considered as a constant, every emitting point of the chip is supposed to be radiating the same flux. Equation (15) can be very practical because an analytical representation of $I_{in}(\theta_p,\varphi_p)$ can be accurately obtained for almost any LED [28], including new LEDs with designed radiation pattern [29,30].

Eqs. (5) and (15) are quasi-radiometric equations of the LTE that explicitly depend on the optical and structural parameters of both the LED chip and the encapsulating lens. These equations can be easily evaluated by any mathematical software for different geometries involving a spherical package. Several examples are presented in section 5, which are verified with Monte-Carlo ray-trace simulation using ASAP optical software. Before that, let us analyze the case of a flat encapsulation.

## 4. Special case of a flat package

Eqs. (5) and (15) should be reduced to the LTE of a flat encapsulant if $R \to \infty$, but the calculation of $\eta$ becomes impractical because the coordinate system is moved to infinity. To solve this problem, we note that the area segment of the package that emits a bunch of rays in the same direction remains constant for every angle θ (see Fig. 2). This simplifies the calculation of the LTE of a flat package by using Eq. (3). In this case, when the flat encapsulant is thin, i.e. $h^2/l^2 < (n_{out}/n_{in})^2 - 1$, and without considering multiple internal reflections, Eq. (3) reduces to

$$\phi_{out} = \frac{n_{out}^2}{n_{in}^2} \int\int_{S_{in}} \int_0^{2\pi} \int_0^{\pi/2} L_{in}(\theta_p, \varphi_p, S) T(\theta) \cos\theta \sin\theta \, d\theta \, d\varphi \, dS, \qquad (16)$$

Let us now make two simplifications that illustrate practical cases. First, if only the angular intensity distribution is available, Eq. (16) can be approached by

$$\phi_{out} = \frac{n_{out}^2}{n_{in}^2} \int\int_{S_{in}} \int_0^{2\pi} \int_0^{\pi/2} I_{in}(\theta_p, \varphi_p) f(S) T(\theta) \left(\frac{\cos\theta}{\cos\theta_p}\right) \sin\theta \, d\theta \, d\varphi \, dS, \qquad (17)$$

where $\theta$ is shown in Fig 2, $\varphi$ is the azimuthal angle (we note that $\varphi_p = \varphi$), and $T$ is given by Eq. (10). This approach states that $T$ is not function of $S$. Considering this, and rearranging the integral of $\phi_{out}$ and $\phi_{in}$, the η equation can be simplified to

$$\eta = \frac{n_{out}^2}{n_{in}^2} \frac{\int_0^{\pi/2} \int_0^{\pi/2} I(\theta_p, \varphi) T(\theta) \left(\frac{\cos\theta}{\cos\theta_p}\right) \sin\theta \, d\theta \, d\varphi}{\int_0^{\pi/2} \int_0^{\pi/2} I(\theta, \varphi) \sin\theta \, d\theta \, d\varphi}, \qquad (18)$$

The other case is when the die is a Lambertian source, $L_{in}$ =constant. The efficiency simply becomes

$$\eta = \frac{2n_{out}^2}{n_{in}^2} \int_0^{\pi/2} T(\theta) \cos\theta \sin\theta \, d\theta = \frac{2n_{out}^2}{n_{in}^2} \overline{T}. \qquad (19)$$

Eq. (19) is well known, and $\overline{T}$ is called the angle averaged transmission coefficient, [31].

## 5.  EXAMPLES

In this section we apply Eqs. (5), (15), (18) and (19) for several different cases. The results are compared with those obtained through Monte-Carlo ray tracing with one million rays. As already stated, we do not consider multiple reflections inside the encapsulant, and then $\eta$ represents the minimum efficiency that may be obtained. In the first three examples, we consider the chip surfaces to be Lambertian sources, i.e., we consider the radiance $L_{in}$ to be a constant. In the last examples we evaluate $\eta$ for a chip with emitting distribution in function of $(\theta_p,\varphi_p)$ [28]. In particular, we consider both photonic quasi crystal and photonic crystal LEDs.

For a square chip the differential of area *dS*, is *dxdy* for the large surface and *dxdz* or *dydz* for the lateral chip faces. In the following examples each chip is square, but other shapes can be easily analyzed. For example, a triangular chip shape or others can perform more efficient than the square shape [32,33], for those cases *dS* can also be easily selected.

### *4.A. Flat Chip*

First, consider the simple case of an LED with a flat chip inside a spherical package as shown in Fig. 3. For this case we obtain $\eta$ varying: in Fig. 3(a) the encapsulant radius $R$, in Fig. 3(b) the $z$ position of the chip, and the refractive index $n$ in Fig. 3(c).

Concerning the package dimensions issue, Fig. 3(a) shows how $\eta$ considerable decreases when the package radius is similar to the chip size. In this case light recycling becomes important. It can be shown that if one considers light recycling in LED chips, then the extracted light increases with respect to the case in which is disregarded [15,16]. Light recycling can also considerably increase the brightness of LEDs [34,35]. This denotes the importance of both an analysis of light recycling, and a good design of the mirror cup.

Fig. 3(b) illustrates a package with the chip located in the optical axis, but outside the curvature center. This can be a manufacturing error or can be deliberately produced by new packaging methods [4]. But $\eta$ considerable decreases when the chip position is larger than -0.6R. Nevertheless, a recent design locates the chip in the bottom of a spherical package (i.e. z~-0.9R), but the efficiency is increased by a scattering mirror that breaks the TIR, and thus sends the rays in the forward direction [11].

It is well known that the chip extraction efficiency increases with the refractive index $n$ of the surrounding encapsulation. However, Fig. 3(c) shows how $\eta$ decreases when $n$ increases. Although small, this tradeoff between chip extraction efficiency and package transmission efficiency affects the overall LED efficiency, but it usually is not commented in the literature.

## 4.B. Four flat chips in the central plane ($z=0$)

To achieve a high flux level required in many lighting applications, several chips are integrated into a single package [27]. Moreover, manufacturing of LEDs with multiple chips has increased over the last few years due to increments in the thermal efficiency [36-38].

In a multichip arrangement, there are many possible geometries. For simplicity, we choose to address four identical chips located in the package central plane and symmetrically positioned, as seen in Fig. 4. We calculate the $\eta$ in function of the chip separation, compared, like in the previous cases, with a ray trace simulation. The results are also shown in Fig. 4. It can be observed that $\eta$ begins to decrease considerably when the chip separation is $0.6R$, which is a similar number as in Fig. 3(b).

## *4.C. Chip with lateral faces*

Recent efforts to enhance optical efficiency include the implementation of a spherical package around a suspended LED chip [10]. We illustrate this type of package by modeling a square chip with two main faces and four lateral faces, see Fig. 5. We consider that the emission proportion is 40% for the main faces and 15% for each of the lateral faces [11]. These values may vary from LED to LED, i.e. as the chip area and the active layer absorption are increased, the proportion of light emitted from lateral faces decreases [13,39]. In this simulation, the emission is Lambertian in all faces. Besides, the proportion of main face size and lateral face size is one tenth, that is, $a = 10c$. The simulation is shown in Fig. 5.

## *4.D. Chip with photonic crystal*

To exemplify Eq. (15), let us consider both photonic quasi crystal (PQC) and photonic crystal (QC) LEDs. PQCs and PCs structures in close proximity to the emitting region promise considerable increase in light extraction [14-17,40-44], and additional properties such as beam shaping into specific radiation [14,40-43]. Here we calculate $\eta$ by using Eq. (15). We simulate one photonic crystal organic LED and three different PQC LEDs with different angular intensity distributions. Some of these radiation patterns are measured and other are simulated distributions [42-44]. PQC LEDs have an isotropic beam profile in the azimuthal direction, then for practical purposes $I_{in}(\theta_p,\varphi_p)=I(\theta_p)$. We analytically model the normalized radiation pattern $I(\theta_p)$ of PQC LEDs by using a sum of 2 or 3 Gaussian functions [28]:

$$I(\theta_p) = \sum_m g_{m,1} \exp\left(-\ln 2 \left(\frac{|\theta_p| - g_{m,2}}{g_{m,3}}\right)^2\right). \tag{20}$$

However, in PC LEDs the in-plane angle $\varphi_p$ plays a critical role. This anisotropy in the radiation pattern depends on the lattice type, e.g. a PC LED with triangular lattice has a 6-fold symmetry. This nonuniformity with azimuthal angles can be sophisticated [39], but some are easier to model [14,45]. In the latter case, $I_{in}(\theta_p,\varphi_p)$ may be approximated by:

$$I(\theta_p,\varphi_p) = \sum_m G(\varphi_p)_{m,1} \exp\left(-\ln 2 \left(\frac{|\theta_p| - G(\varphi_p)_{m,2}}{G(\varphi_p)_{m,3}}\right)^2\right), \tag{21}$$

where

$$G(\varphi_p)_{mn} = g1_{mn} \cos^2(N\varphi_p) + g2_{mn} \sin^2(N\varphi_p)$$

Here $g1_{mn}$, $g2_{mn}$ and $N$ are constants that mainly depend on the PC structural parameters. The two sets of coefficients $g1_{mn}$ and $g2_{mn}$ model the radiation pattern along the two main directions of the PC, e.g. these are the ΓM and ΓK directions of a triangular lattice [14], and for a square lattice are the ΓM and ΓX directions. The azimuthal symmetry is introduced by $N$, e.g. $N=3$ for a triangular lattice, and for a square lattice $N=2$.

Fig. 6 shows $\eta$ in function of the chip size for the package examined in Fig. 3(a). We analyze the chip size dependence because the high mode coupling makes photonic crystal chips suitable for being large. Inset figures show the radiation pattern. The coefficients of Eqs. (20) and (21) that we used in simulations of Fig. 6 are shown in Table 1. These LED radiation patterns are measured in [44] for PC (4-fold symmetry, $N=2$) and in [43] for PQC-1, and are simulated in [42] for PQC-2 and PQC-3.

From Fig. 6, it can be observed that $\eta$ begins to decrease considerably when the chip is larger than $0.5R$ independently of the radiation pattern shape. In addition, we observed that, as the number of light rays increases the ray trace simulation approaches more and more the analytic calculation, which is more pronounced in Figs. 6(c) and 6(d).

Note that the light pattern near a PC chip can be quite different from the far-field approach given by Eqs. (20) and (21). Moreover, wave effects could significantly reach the encapsulant near the chip corners [46,47], and then the classic radiometric approach could not be precise. Therefore, this example and the next are approximations when the encapsulant is very small ($R\rightarrow 0.7a$ for a square chip). This problem opens a window for further studies. As was done for encapsulated LEDs and LED arrays [21,48], it could be very useful to find a far-field condition for PC LED chips. For example, to give the distance beyond which leaky modes continuum would not reach the encapsulant boundaries [46].

In addition, further work could investigate the LTE optimization of a package with PC LEDs, as made for the light extraction efficiency of PC LED chips [14], considering all structural parameters of interest, e.g. $n_{out}/n_{in}$ and $z$-position of chip.

*4.E. Flat Package with photonic crystal*

Fig. 7(a) shows the efficiency in function of the refractive index $n_{in}=n$ for the photonic crystal LEDs of Table 1 (by using Eq. (18)). We analyze this case because embedding a PC LED in a flat package with a different index of refraction (refractive index contrast) changes the optical behavior of the device [39,40]. From Fig. 7(a), it is evident that the LTE is smaller if the LED radiation pattern is wider, i.e. it is larger if the angular intensity distribution of the LED chip is more directional. We illustrate this behavior in Fig. 7(b). For simplicity, we use a simple cosine-power radiation pattern, with half-intensity viewing angles (half width half maximum angle) ranging from 10° to 60°.

## 6. CONCLUSIONS

Practically all LEDs are packaged within a transparent epoxy or silicone-gel. There are many types of packages, but spherical and flat are the most popular. But even for a spherical package, there are many possible structures; just consider a multi chip LED, or the wide variety of LED dies. In this context, a realistic equation for the LTE of a spherical package may provide useful assistance in the optical analysis of LEDs. From radiation transfer theory, when trying to calculate the LTE of an LED spherical package, we arrived to very complicated transcendental equations for the integration limits of the radiometric equation, thereby making the computation intractable. We overcame this problem by developing a quasi-radiometric equation, which incorporates some concepts of Monte-Carlo ray tracing into the context of radiometry. This approach includes the extended source nature of the LED chip, and the chip radiance distribution. The $\eta$ equation is an explicit function of key parameters such as chip geometry, chip radiance, chip location, package size, and refractive index. To our knowledge, a generalized analysis like this has not been reported before. We analyzed several packaging configurations of interest, e.g., a hemispherical lens with multiple chips, and a package with a PC LED and with PQC LED. We compared the analytic calculation with a Monte-Carlo ray-tracing simulation, obtaining almost identical values. Additional applications include the calculation of the LTE of LED encapsulations with unconventional phosphor layer geometries [49,50], and non-square LED chips [32,33]. Further research of this analysis include: the incorporation of light recycling and the mirror cup in the equations; a study of LTE optimization; and the development of near field condition for determining when to use ray-tracing or a rigorous electromagnetic approach.

## ACKNOWLEDGMENTS

The authors would like to thank the anonymous reviewer one for his valuable comments that improved the manuscript and opened new frontiers for further research.

**List of Figure Captions**

Fig. 1. Geometry of an LED with spherical package of radius $R$.

Fig.2. Geometry of an LED with flat package.

Fig. 3. Results obtained for a square central chip inside a spherical package. In (a) and (b) $n=1.5$, and in (b) and (c) $R=10a$.

Fig. 4. Results for the four identical chips in the central plane of a spherical encapsulant in function of the total emitting area in both analytic method and ray trace simulation. In this case $z=0$, $n=1.5$ and $R=10a$.

Fig. 5. Efficiency varying the encapsulant radius for the chip with lateral faces. For this geometry $n=1.5$, $z=0$, and $a=10c$.

Fig. 6. Efficiency varying the size of the LED chip, for several LED radiation patterns. Results obtained for a square central chip inside a spherical package with $n=1.5$ and $z=0$, see inset of Fig. 3(a).

Fig.7. Flat package. (a) Efficiency varying the index of refraction of package for several LED radiation patterns (see insets of Fig. 6) with $n_{out}=1$, and $n_{in}=n$. (b) LTE of directional LEDs for several half-angles $\theta_{1/2}$.

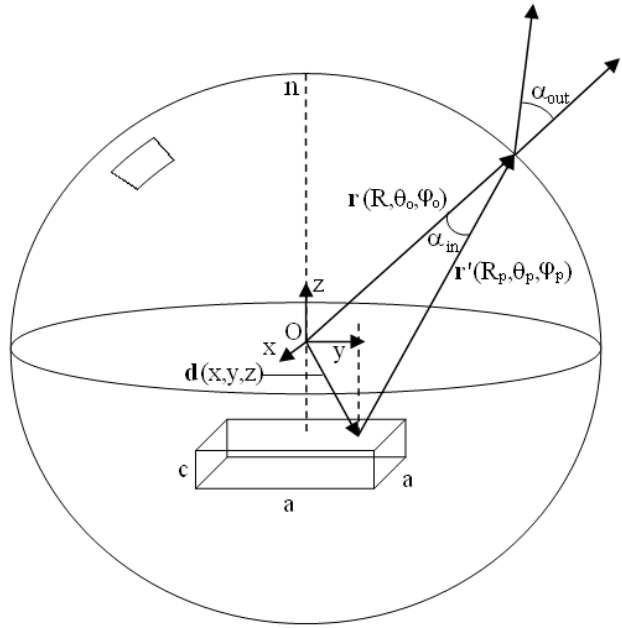

Fig. 1. Geometry of an LED with spherical package of radius *R*.

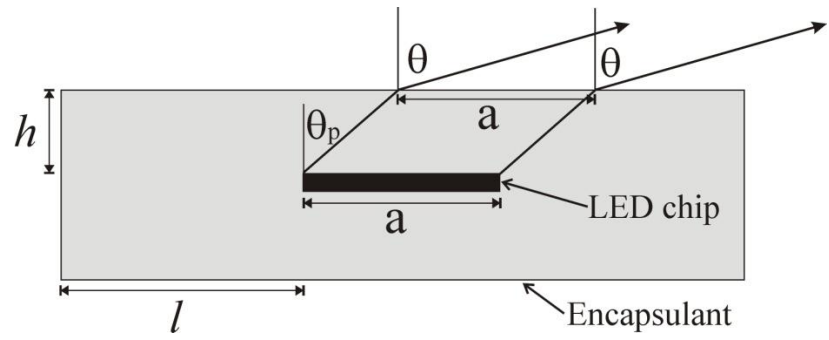

Fig.2. Geometry of an LED with flat package.

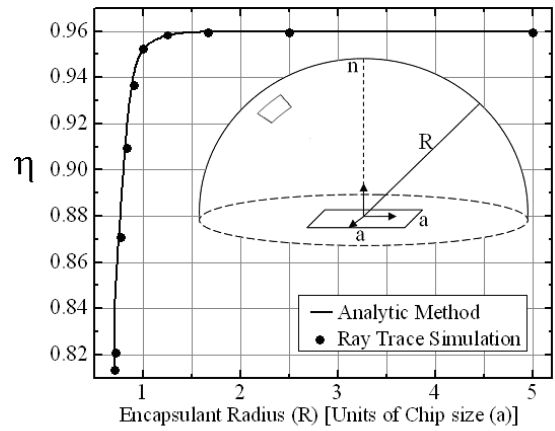

(a)

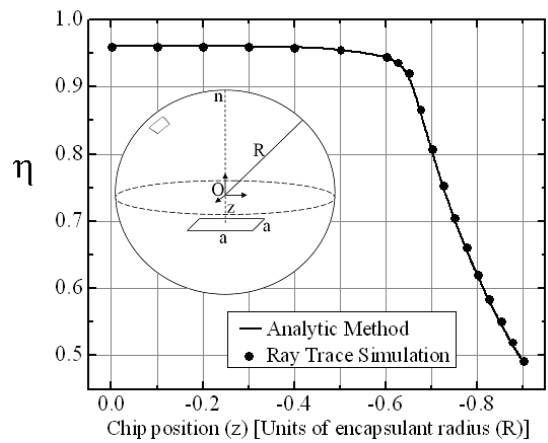

(b)

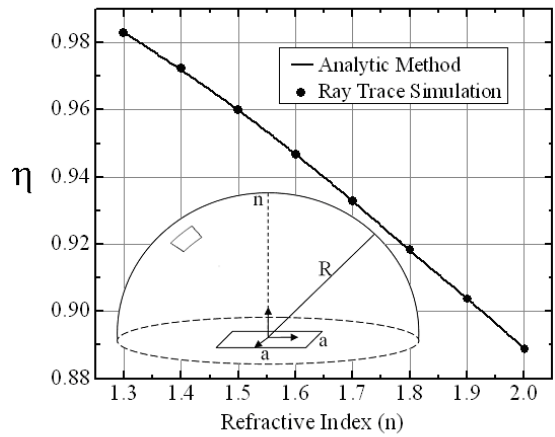

(c)

Fig. 3. Results obtained for a square central chip inside a spherical package. In (a) and (b) $n = 1.5$, and in (b) and (c) $R = 10a$.

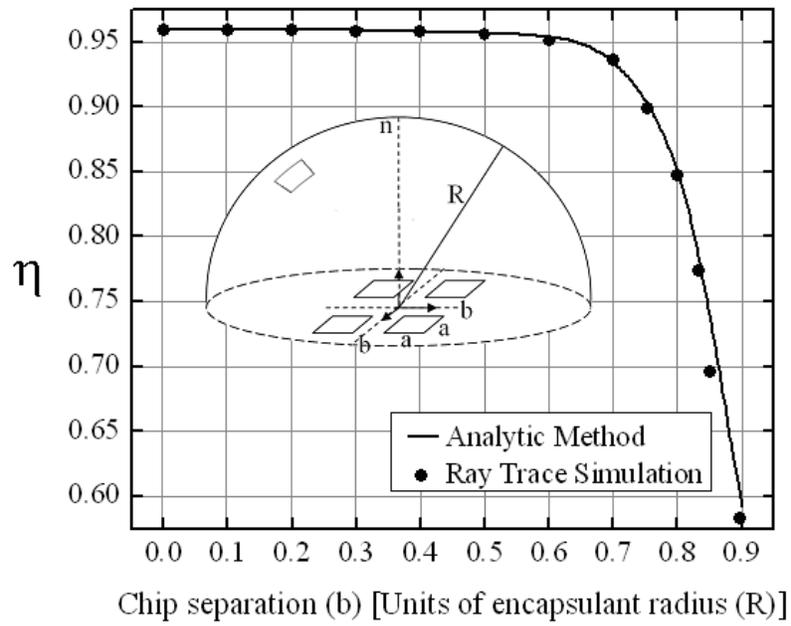

Fig. 4. Results for the four identical chips in the central plane of a spherical encapsulant in function of the total emitting area in both analytic method and ray trace simulation. In this case $z=0$, $n=1.5$ and $R=10a$.

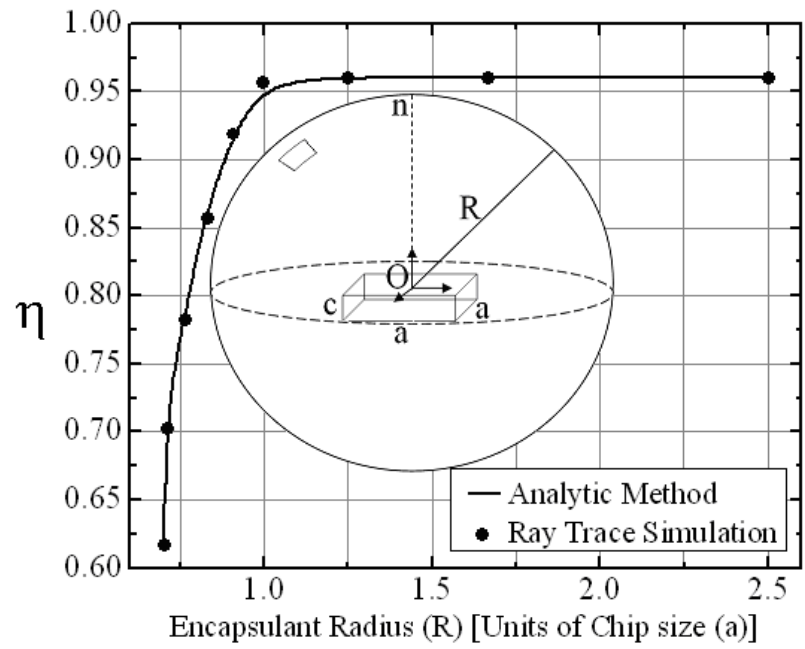

Fig.5. Efficiency varying the encapsulant radius for the chip with lateral faces. For this geometry $n=1.5$, $z=0$, and $a=10c$.

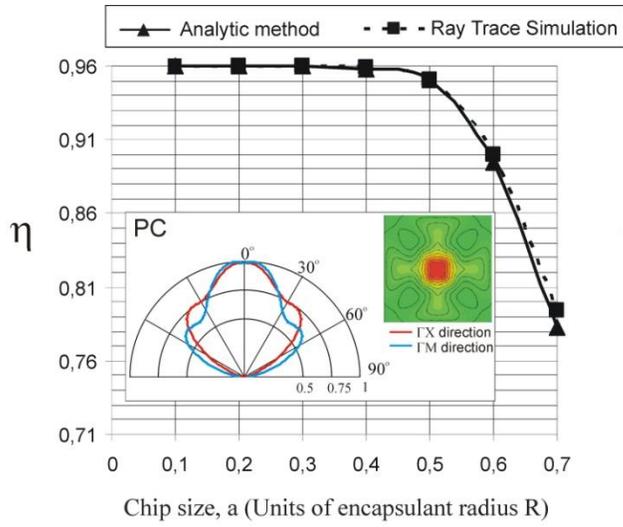
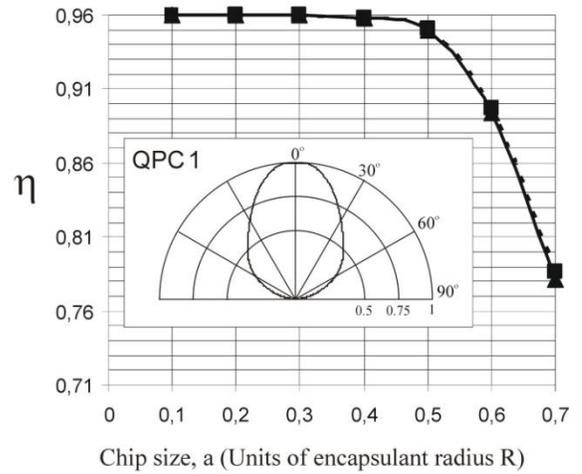
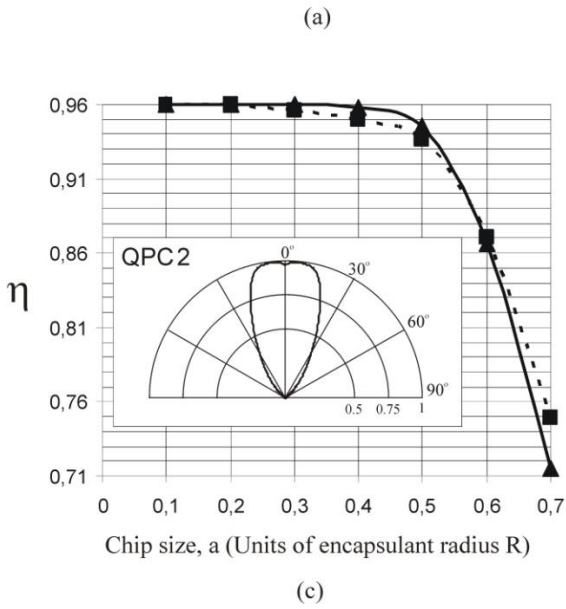
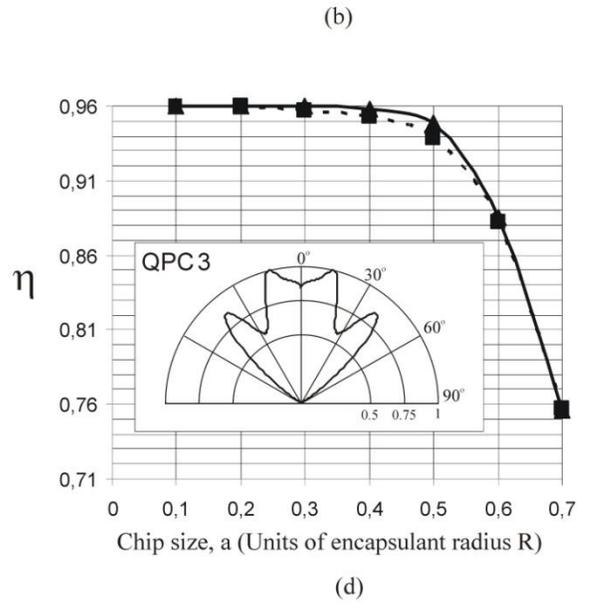

Fig.6. Efficiency varying the size of the LED chip, for several LED radiation patterns. Results obtained for a square central chip inside a spherical package with $n=1.5$ and $z=0$, see inset of Fig. 3(a)

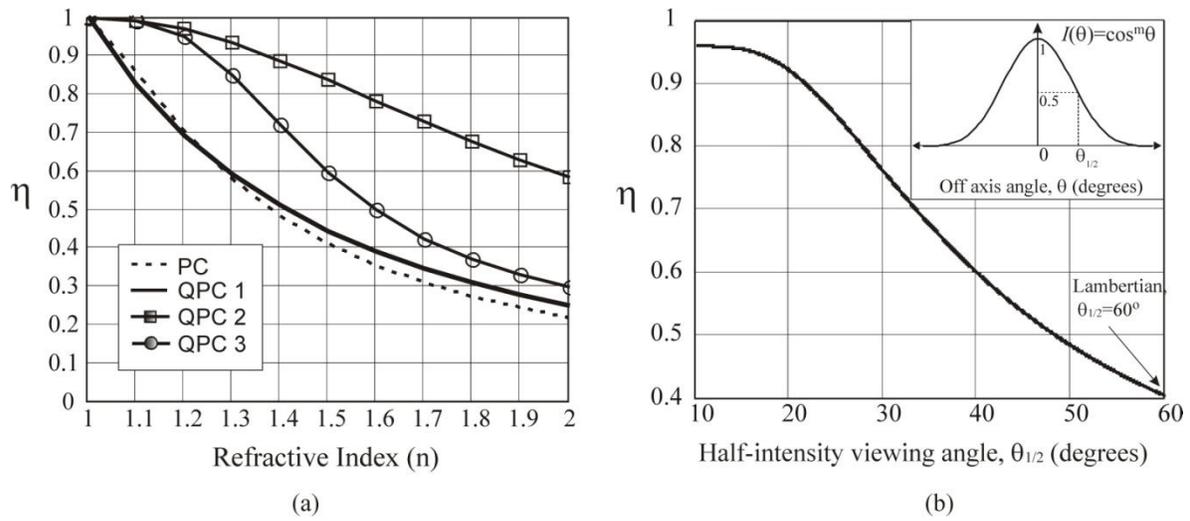

Fig.7. Flat package. (a) Efficiency varying the index of refraction of package for several LED radiation patterns (see insets of Fig. 6) with $n_{out}=1$, and $n_{in}=n$. (b) LTE of directional LEDs for several half-angles $\theta_{1/2}$.

Table 1. Coefficients of Eqs. (20) and (21) used in calculations of Figs. 6 and 7.

| Coefficient | PC $g1/g2$ | QPC 1 | QFC 2 | QFC 3 |
|---|---|---|---|---|
| $g_{1,1}$ | 1.028/ 1.045 | 0.993 | 0.993 | 0.899 |
| $g_{2,1}$ | 0.576/ 0.522 | 0.298 | 0.095 | 0.191 |
| $g_{3,1}$ | | | | 0.704 |
| $g_{1,2}$ | 1.338/ 4.00 | 1.882 | 3.418 | 5.958 |
| $g_{2,2}$ | 45.16/ 56.00 | 53.06 | 15.42 | 15.09 |
| $g_{3,2}$ | | | | 41.62 |
| $g_{1,3}$ | 26.55/ 29.139 | 29.681 | 22.862 | 21.68 |
| $g_{2,3}$ | 16.85/ 16.65 | 21.655 | 5.828 | 4.145 |
| $g_{3,3}$ | | | | 8.842 |